\documentclass[aps,pre,twocolumn,floatfix,showpacs]{revtex4}
\usepackage{epsf,amsmath,amssymb,verbatim}
\begin{document}
\newcommand{\be}{\begin{equation}}
\newcommand{\ee}{\end{equation}}
\newcommand{\kin}{k_{\rm in}}
\newcommand{\kout}{k_{\rm out}}
\title{Internet data packet transport: from global topology
to local queueing dynamics}
\author{H.~K. Lee$^1$, K.-I. Goh$^2$, B. Kahng$^{3,4}$, and D. Kim$^3$}
\affiliation{{$^1$ School of Physics, Korea Institute for Advanced
Study, Seoul 130-722, Korea}\\
{$^2${Center for Cancer Systems Biology, Dana-Farber Cancer
Institute, Harvard Medical School, Boston, MA 02115 and}\\
{Center for Complex Network Research and Department of Physics,
University of Notre Dame, Notre Dame, IN 46556}}\\
{$^3$ School of Physics and Center for Theoretical Physics, Seoul
National University NS50, Seoul 151-747, Korea}\\ {$^4$Center for
Nonlinear Studies, Los Alamos National Laboratory, Los Alamos, NM
87545}}
\date{\today}

\begin{abstract}
We study structural feature and evolution of the Internet at the
autonomous systems level. Extracting relevant parameters for the
growth dynamics of the Internet topology, we construct a toy model
for the Internet evolution, which includes the ingredients of
multiplicative stochastic evolution of nodes and edges and
adaptive rewiring of edges. The model reproduces successfully
structural features of the Internet at a fundamental level. We
also introduce a quantity called the load as the capacity of node
needed for handling the communication traffic and study its
time-dependent behavior at the hubs across years. The load at hub
increases with network size $N$ as $\sim N^{1.8}$. Finally, we
study data packet traffic in the microscopic scale. The average
delay time of data packets in a queueing system is calculated, in
particular, when the number of arrival channels is scale-free. We
show that when the number of arriving data packets follows a power
law distribution, $\sim n^{-\lambda}$, the queue length
distribution decays as $n^{1-\lambda}$ and the average delay time
at the hub diverges as $\sim N^{(3-\lambda)/(\gamma-1)}$ in the $N
\to \infty$ limit when $2 < \lambda < 3$, $\gamma$ being the
network degree exponent.
\end{abstract}
 \maketitle In recent years,
the Internet has become one of the most influential media in our
daily life, going beyond in its role as the basic infrastructure
in this technological world. Explosive growth in the number of
users and hence the amount of traffic poses a number of problems
which are not only important in practice for, e.g.,  maintaining
it free from any undesired congestion and malfunctioning, but also
of theoretical interests as an interdisciplinary
topic~\cite{internetbook}. Such interests, also stimulated by
other disciplines like biology, sociology, and statistical
physics, have blossomed into a broader framework of network
science~\cite{rmp,portobook,siam,report}. In this Letter, we first
review briefly previous studies of Internet topology and the data
packet transport on global scale, and next study the delivery
process in queueing system of each node embedded in the Internet.

The Internet is a primary example of complex networks. It consists
of a large number of very heterogeneous units interconnected with
various connection bandwidths, however, it is neither regular nor
completely random. In their landmark paper, Faloutsos {\it et
al.}\ \cite{fal3} showed that the Internet at the autonomous
systems (ASes) level is a scale-free (SF) network \cite{physica},
meaning that degree $k$, the number of connections a node has,
follows a power-law distribution,
\begin{equation}
P_d(k)\sim k^{-\gamma}. \label{eq:degree}
\end{equation}
The degree exponent $\gamma$ is subsequently measured and
confirmed in a number of studies to be $\gamma\approx 2.1(1)$. The
power-law degree distribution implies the presence of a few nodes
having a large number of connections, called hubs, while most
other nodes have a few number of connections.

It is known that the degrees of the two nodes located at each end
of a link are correlated each other. As the first step, the
degree-degree correlation can be quantified in terms of the mean
degree of the neighbors of a given node with degree $k$ as a
function of $k$, denoted by $\langle k_{\rm nn} \rangle (k)$
\cite{vespig}, which behaves in another power law as
\begin{equation}
\langle k_{\rm nn}\rangle (k) \sim k^{-\nu}. \label{eq:knn_power}
\end{equation}
For the Internet, it decays with $\nu\approx 0.5$ measured from
the real-world Internet data \cite{routeviews,nlanr}.

The Internet has modules within it. Such modular structures arise
due to regional control systems, and often form in a hierarchical
way \cite{maslov_PRL}. Recently, it was argued that such modular
and hierarchical structures can be described in terms of the
clustering coefficient. Let $C_i$ be the local clustering
coefficient of a node $i$, defined as $C_i=2e_i/k_i(k_i-1)$, where
$e_i$ is the number of links present among the neighbors of node
$i$, out of its maximum possible number $k_i(k_i-1)/2$. The
clustering coefficient of a network, $C$, is the average of $C_i$
over all nodes. $C(k)$ means the clustering function of a node
with degree $k$, i.e., $C_i$ averaged over nodes with degree $k$.
When a network is modular and hierarchical, the clustering
function follows a power law, $C(k) \sim k^{-\beta}$ for large
$k$, and $C$ is independent of system size
$N$~\cite{Ravasz02,Ravasz03}. For the Internet, it was measured
that the clustering coefficient is $C_{\rm AS}\approx 0.25$ and
the exponent $\beta\approx 0.75$~\cite{vpsv}.

\begin{figure}[t]
\centerline{\epsfxsize=8cm \epsfbox{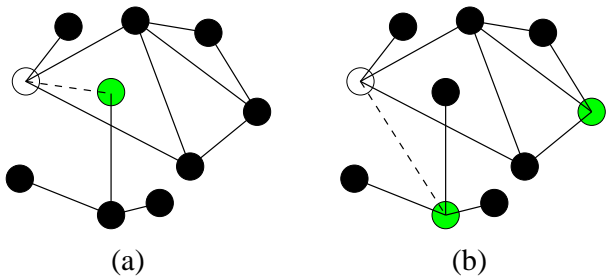}} \caption{Shown is
the adaptive rewiring rule. A node (white) detaches one of its
links from a node (green or gray) in (a), and attaches it to one
of the nodes (green or gray) with degree 3, larger than 2 of the
detached node, in (b).} \label{adap}
\end{figure}

There are many known models to mimic the Internet topology. Here
we introduce our stochastic model evolving through the following
four rules. This model is based on the model proposed by Huberman
and Adamic~\cite{ha}, which is a generic model to reproduce a
uncorrelated SF network and we modify it by adding the adaptation
rule~\cite{fluc}, which results in generating the degree-degree
correlations. The rules are as follows: {(i)} {\em Geometrical
growth}: At time step $t$, geometrically increased number of new
nodes, $\alpha N(t-1)$, are introduced in the system with the
empirical value of $\alpha=0.029$. Then following the empirical
fact $\langle k_{\rm new}\rangle_t \approx 1.34$, each of newly
added nodes connects to one or two existing nodes according to the
preferential attachment (PA) rule~\cite{ba}. {(ii)} {\em
Accelerated growth}: Each existing node increases its degree by
the factor empirical value of $\approx 0.035$. These new internal
links are also connected following the PA rule. {(iii)} {\em
Fluctuations}: Each node disconnects existing links randomly or
connects new links following the PA rule with equal probability.
The variance of this noise is given as $\sigma^2\approx (0.14)^2$
measured from empirical data. {(iv)} {\em Adaptation}: When
connecting in step (iii), the PA rule is applied only within the
subset of the existing nodes consisting of those having larger
degree than the one previously disconnected. This last constraint
accounts for the adaptation process. The adaptive rewiring rule is
depicted in Fig.~\ref{adap}.

Through this adaptation model, we can reproduce generic features
of the Internet topologies successfully which are as follows:
First, the degree exponent is measured to be $\gamma_{\rm model}
\approx 2.2$, close to the empirical result $\gamma_{\rm AS}
\approx 2.1(1)$. Second, the clustering coefficient is measured to
be $C_{\rm model} \approx 0.15(7)$, comparable to the empirical
value $C_{\rm AS} \approx 0.25$. Note that without the adaptation
rule, we only get $C\approx 0.01(1)$. The clustering function
$C(k)$ also behaves similarly to that of the real-world Internet,
specifically, decaying in a power law with $\beta\approx 1.1(3)$
roughly for large $k$~\cite{book1}, but the overall curve shifts
upward and the constant behavior for small $k$ appears. Third, the
mean degree function $\langle k_{\rm nn}\rangle (k)$ also behaves
similarly to that of the real-world Internet network, but it also
shifts upward overall. In short, the behaviors of $C(k)$ and
$\langle k_{\rm nn} \rangle(k)$ of the adaptation model are close
to those of the real Internet AS map, but with some discrepancies
described above. On the other hand, recently another toy
model~\cite{serrano05} has been introduced to represent the
evolution of the Internet topology. The model is similar to our
model in the perspective of including the multiplicative
stochastic evolution of nodes and edges as well as adaptive
rewiring of edges. However, the rewiring dynamics is carried out
with the incorporation of user population instead of degree of
node we used here.

\begin{figure}[t]
\centerline{\epsfxsize=8cm \epsfbox{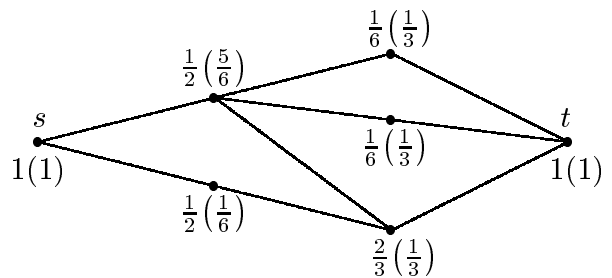}} \caption{The load
at each node due to a unit packet transfer from the node $s$ to
the node $t$, $\ell_i^{s\rightarrow t}$. In this diagram, only the
nodes along the shortest paths between $(s,t)$ are shown. The
quantity in parentheses is the corresponding value of the load due
to the packet from $t$ to $s$, $\ell_i^{t\rightarrow s}$.}
\label{load_def}
\end{figure}

Next, we study the transport of data packet on the Internet. Data
packets are sent and received over it constantly, causing
momentary local congestion from time to time. To avoid such
undesired congestion, the capacity, or the bandwidth, of the
routers should be as large as it can handle the traffic. First we
introduce a rough measure of such capacity, called the load and
denoted as $\ell$~\cite{load}. One assumes that every node sends a
unit packet to everyone else in unit time and the packets are
transferred from the source to the target only along the shortest
paths between them, and divided evenly upon encountering any
branching point. To be precise, let $\ell_i^{s\rightarrow t}$ be
the amount of packet sent from $s$ (source) to $t$ (target) that
passes through the node $i$ (see Fig.~\ref{load_def}). Then the
load of a node $i$, $\ell_i$, is the accumulated sum of
$\ell_i^{s\rightarrow t}$ for all $s$ and $t$, $\ell_i =
\sum_{s\neq t} \ell_i^{s\rightarrow t}$. In other words, the load
of a node $i$ gives us the information how much the capacity of
the node should be in order to maintain the whole network in a
free-flow state. However, due to local fluctuation effect of the
concentration of data packets, the traffic could be congested even
for the capacity of each node being taken as its load. The
distribution of the load reflects the high level of heterogeneity
of the Internet: It also follows a power law,
\begin{equation}
P_{l}(\ell)\sim \ell^{-\delta}, \label{eq:load}
\end{equation}
with the load exponent $\delta\approx 2.0$ for the Internet. For
comparison, the quantity ``load" is different from the
``betweenness centrality"~\cite{freeman} in its definition. In
load, when a unit packet encounters a branching point along the
shortest pathways, it is divided evenly with the local information
of branching number, while in betweenness centrality, it can be
divided unevenly with the global information of the total number
of shortest pathways between a given source and
target~\cite{book2}. Despite such a difference, we find no
appreciable difference in practice for the numerical values of the
load and the betweenness centrality for a given network.

The load of a node is highly correlated with its degree. This
suggests a scaling relation between the load and the degree of a
node as $\ell \sim k^{\eta}$ and the scaling exponent $\eta$ is
estimated as $\eta=1.06\pm 0.03$ for January 2000 AS
map~\cite{vpsv,book1}. In fact, if one assumes that the ranks of
each node for the degree and the load are the same, then one can
show that the exponent $\eta$ depends on $\gamma$ and $\delta$ as
$\eta= (\gamma-1)/(\delta-1)$ with $\gamma\approx 2.1$ and
$\delta\approx 2.0$, and we have $\eta\approx 1.1$, which is
consistent with the direct measurement.

\begin{figure}
\centerline{\epsfxsize=7cm \epsfbox{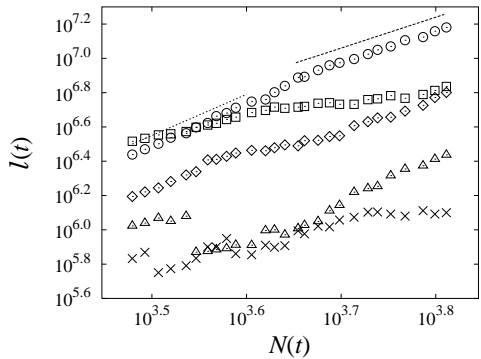}} \caption{ Time
evolution of the load versus $N(t)$ at the ASes of degree-rank
1({\Large$\circ$}), 2 ($\Box$), 3 ($\Diamond$), 4 ($\triangle$),
and 5 ($\times$). The dashed line for larger $N$ has slope 1.8,
drawn for the eye. } \label{load_hub}
\end{figure}

The time evolution of the load at each AS is also of interest.
Practically, how the load scales with the total number of ASes
(the size of the AS map) is an important information for the
network management. In Fig.~\ref{load_hub}, we show $\ell_i(t)$
versus $N(t)$ for 5 ASes with the highest rank in degree, i.e., 5
ASes that have largest degrees at $t=0$. The data of
$\{\ell_i(t)\}$ shows large fluctuations in time. Interestingly,
the fluctuation is moderate for the hub, implying that the
connections of the hub is rather stable. The load at the hub is
found to scale with $N(t)$ as $\ell_h(t)\sim N(t)^{\mu}$, but the
scaling shows a crossover from $\mu\approx 2.4$ to $\mu\approx1.8$
around $t\approx14$.

Internet traffic along the shortest pathways yields inconvenient
queue congestions at hubs in SF networks. Many alternative routing
strategies have been introduced to reduce the load at hub and
improve the critical density of the number of packets displaying
the transition from free-flow to congested
state~\cite{tadic04,rodgers,arenas,holme,echenique,greiner,duch,zoltan}.

Transport of data packets also relies on queueing process of an
individual AS. Here we extend existing queueing theory~\cite{qt}
to the case where arrival channels are multiple, in particular,
when their number distribution follows a power law, aiming at
understanding the transport in SF networks. For simplicity, we
assume that the arrival and processing rates of an individual
channel are the same, and they are independent of degree of a
given AS. Time is discretized and unit time is given as the
inverse of the rate.

\begin{figure}
\centerline{\epsfxsize=7cm \epsfbox{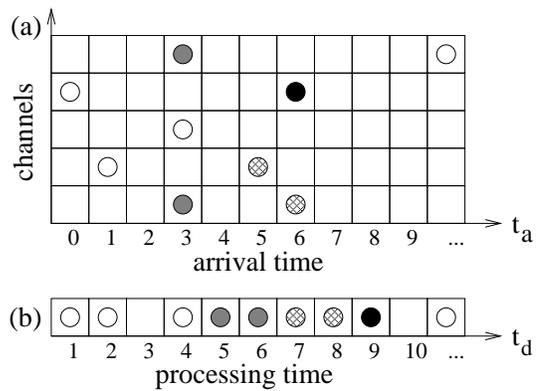}} \caption{(a)
Snapshot inside buffer with arriving packets. Each row represents
a communication channel, and circles therein are the sequence of
incoming packets. The integers on the horizontal axis indicate
arriving time-steps of each packet. Open circles stand for packets
not delayed. Packets delayed are represented by three kinds of
filled circles according to their own delaying mechanism. See text
for details. The consequent delivery sequence is shown in (b) with
processing time-step.} \label{modeledRouting}
\end{figure}

Delay of packet delivery in our queueing process originates from
two sources. For the one, owing to multiple arriving channels,
multiple packets can arrive at a given queueing system in a unit
time interval, and are accumulated in the buffer. For example,
grey circles in Fig.~\ref{modeledRouting} represent such a case.
This type of delay is referred to as the delay type 1 (DT1) below.
For the other, the delay is caused by preceding packets in the
buffer, which can happen under the first-in-first-out rule. The
hatched circles in Fig.~\ref{modeledRouting} demonstrate this
case. This case is referred to as the delay type 2 (DT2). Then any
delay can be decomposed into the two types. The black circle in
Fig.4 is such a packet, delayed by both DT1 and DT2. We calculate
the average delay time for each type, separately, and combine them
next.

To proceed, we first define $p_n$ as the probability that $n$
packets arrive at a given queueing system at the same time. For
the DT1 case, if $q_m$ denotes the probability that a packet is
delayed $m$ time steps, we find
\begin{equation}
q_m= p_0\delta_{0,m} + \sum_{n=m+1}^{\infty} {p_n \over n},
\label{qm}
\end{equation}
where $\delta_{i,j}$ is the Kronecker delta function. Then, the
average of delay time steps through the DT1 process is obtained as
\begin{equation}
\langle m \rangle_q = \sum_{n=2}^{\infty} {p_n \over n}
\sum_{m=1}^{n-1} m ={ {\langle n \rangle_p -1+ p_0}\over 2} ,
\label{av_r}
\end{equation}
where $\langle \cdots \rangle_q$ ($\langle \cdots \rangle_p$) is
the average with respect to the probability $q_m$ ($p_n$).

For the DT2 case, we introduce $r_b(t)$ as the probability that a
packet arrived at time $t$ is delayed $b$ time steps by preceding
delayed packets. In the steady state, we obtain that
\begin{equation}
r_{b^\prime} =  \sum_{b=0}^{{b^\prime}+1} p_{{b^\prime}-b+1}r_b +
p_0r_0\delta_{0,{b^\prime}}. \label{master}
\end{equation}
By using the generating functions ${\cal {R}}(z) \equiv
\sum_{b=0}^{\infty} r_b z^b$ and ${\cal {P}}(z) \equiv
\sum_{n=0}^{\infty} p_n z^n$, we obtain that
\begin{equation}
{\cal {R}}(z)\left[z-{\cal {P}}(z)\right]= p_0r_0(z-1) \ ,
\label{fxp}
\end{equation}
with $p_0 r_0 =1-\langle n \rangle_p$.

The next step is to combine the two types of delays. To this end,
we define $w_\tau$ as the probability that a unit packet is
delayed by $\tau$. Then $w_\tau = \sum_{m=0}^\tau q_{m} r_{\tau
-m}$ since DT1 and DT2 are statistically independent. From this,
the average delay time is obtained as
\begin{equation}
\langle \tau \rangle_w \equiv \sum_{\tau =0}^{\infty} \tau w_\tau
={{\langle n \rangle_p -1+ p_0}\over 2} + {{\langle n^2 \rangle_p
-\langle n \rangle}_p\over {2 \left( 1-\langle n \rangle_p
\right)}} \ .
 \label{averagedelay}
\end{equation}
Thus, a critical congestion occurs when $\langle n \rangle_p=1$,
at which the delay time diverges. The singular behavior in the
form of $(1-\langle n \rangle_p)^{-1}$ was observed numerically in
the study of directed traffic flow in Euclidean
space~\cite{manna}.

We now consider the case where the number of arriving data packets
follows a power law, $p_n \sim n^{-\lambda}$. In fact,
non-uniformity of the number of data packets arriving at a given
node gives rise to self-similar patterns as is well known in
computer science~\cite{walter}. Precise value of the exponent
$\lambda$ has not been reported yet. Moreover, it is not known if
the exponent is universal, independent of bandwidths or degrees in
the SF network. The relation of $\lambda$ to the load exponent
$\delta$, if there is any, is not known either.

If $\lambda<3$, $\langle n^2 \rangle_p$ diverges. For such a
power-law distribution, its generating function ${\cal P}(z)$
develops a singular part and takes the form, when $2 < \lambda
<3$,
\begin{equation}
{\cal {P}}(z)=1-\langle n \rangle_p (1-z)+a(1-z)^{\lambda-1}+
\mathcal{O}\left((1-z)^2 \right),
 \label{s-psi}
\end{equation}
where $a$ is a constant. By using the relation between ${\cal
{P}}(z)$ and ${\cal {R}}(z)$ from Eq.~(\ref{fxp}), we obtain that
\begin{equation}
{\cal {R}}(z)=1-{a\over{1-\langle n \rangle}_p}(1-z)^{\lambda-2}+
\mathcal{O}\left(1-z\right).
 \label{s-chi}
\end{equation}
Therefore, the probability $r_b$ in the delay of the DT2 behaves
as $r_b \sim b^{1-\lambda}$ for large $b$. In other words, the DT2
delay distribution decays slower than that of incoming packets,
$p_n$, and $\langle \tau \rangle_w \sim \langle b \rangle_r$
becomes infinite even when $\langle n \rangle_p <1$.

On the other hand, in real {\em finite} scale-free networks such
as the Internet with the degree exponent $\gamma$, $p_n$ at the
hub has a natural cut-off at $n \sim k_{\rm max} \sim
N^{1/(\gamma-1)}$, in which case we have $\langle n^2 \rangle_p
\sim k_{\rm max}^{3-\lambda}$. Thus from Eq.~(\ref{averagedelay})
the average delay time at the hub scales as
\begin{equation} \langle
\tau \rangle_w \sim k_{\rm max}^{3-\lambda} \sim
N^{(3-\lambda)/(\gamma-1)}
\end{equation} for $2< \lambda < 3$.

In the real-world Internet, the bandwidth of each AS is not
uniform. Nodes with high bandwidth locate at the core of the
network, forming a rich club~\cite{richclub1,richclub2}, however,
their degrees are small. Whereas, nodes with large degree locate
at the periphery of the network with low bandwidth~\cite{hot}.
Therefore, our analysis of the average delay time has to be
generalized incorporating the inhomogeneous bandwidths and arrival
rates \cite{outlook}.

In summary, in the first part of this Letter, we have reviewed the
previous studies of topological properties of the Internet and
introduced a minimal model, the adaptation model to reproduce the
topological properties. Next we studied transport phenomena of
data packets travelling along the shortest pathways from source to
destination nodes in terms of the load. In the second part, we
studied the delivery process of data packets in the queueing
system, in particular, when arrival channels are diverse following
the scale-freeness in the degree distribution.

This work is supported by the KOSEF grants No.
R14-2002-059-01000-0 in the ABRL program.

\end{document}